\documentclass[twocolumn,nofootinbib]{openjournal}
\pdfoutput=1
\usepackage{amsmath}
\usepackage{amssymb}
\usepackage{mathrsfs}
\usepackage{bm}
\usepackage{graphicx}
\usepackage{color}
\newcommand{\beq}{\begin{equation}}
\newcommand{\eeq}{\end{equation}}
\newcommand{\change}[1]{#1}

\def\aap{Astron.\ Astrophys.}
\def\apjs{Astrophys.\ J.\ Supp.}
\def\apjl{Astrophys.\ J.\ Lett.}

\def\mnras{M.N.R.A.S.}
\def\jcap{J.\ Cosmol.\ Astropart.\ Phys}

\begin{document}

\title{Cosmological Inflation in $N$-Dimensional Gaussian Random Fields with Algorithmic Data Compression}
\author{Connor A. Painter and Emory F. Bunn}
\affiliation{Physics Department, University of Richmond, Richmond, VA  23173, USA}

\begin{abstract}
There is considerable interest in inflationary models with multiple inflaton fields. The inflaton field $\boldsymbol\phi$ that has been postulated to drive accelerating expansion in the very early universe has a corresponding potential $V$, the details of which are free parameters of the theory. We consider a natural hypothesis that $V$ ought to be maximally random. We realize this idea by defining $V$ as a Gaussian random field in some number $N$ of dimensions. Given a model that statistically determines the shape of $V$, we repeatedly evolve $\boldsymbol\phi$ under random potentials, cataloging a representative sample of trajectories associated with that model. On anthropic grounds, we impose a minimum with $V=0$ and only consider trajectories that reach that minimum. We simulate each path evolution stepwise through $\boldsymbol\phi$-space while simultaneously computing $V$ and its derivatives along the path via a Gaussian random process. When $N$ is large, this method significantly reduces computational load as compared to methods that generate the potential landscape all at once. Even so, the covariance matrix of constraints on $V$ can quickly become large and cumbersome. To solve this problem, we present data compression algorithms to prioritize the necessary information already simulated, then keep an arbitrarily large portion. With these optimizations, we simulate thousands of trajectories, extract \change{curvature}, tensor, and isocurvature spectra from each, and then assemble statistical predictions of these quantities through repeated trials. We find that the Gaussian random potential is a highly versatile inflationary model with a rich volume of parameter space capable of reproducing modern observations.

\end{abstract}

\maketitle

\section{Introduction} \label{sec:introduction}

Characterization of plausible inflationary models is an extremely high priority in cosmology \citep{dodelson}. Since inflationary theory was developed in the late 1970s and 80s, the cosmological community has sought to constrain the  parameter space of possible inflationary models \citep[e.g.,][]{langlois}. Some parts of this parameter space -- in particular, some choices for the inflaton potential $V$ -- have been explored in exhaustive detail, but the vastness of the parameter space defies complete exploration.

Potentials with a large number of inflaton degrees of freedom $N$ \citep{salopek} are of particular interest, as they may be a natural consequence of string theory \citep{baumann}. Even if we lay string theory aside, it is extremely plausible to imagine that many additional fields may occur as we extend our model of particle physics to the vastly higher energies at which inflation is thought to have occurred. Fully exploring a theory with $N\gg 1$ inflaton fields, with an unknown potential $V(\phi^{(0)},\ldots,\phi^{(N-1)})$, is a computationally daunting task.

Rather than specifying a fixed potential \textit{a priori}, it may be natural to consider models inspired by the ``landscape'' of string theory, in which $V$ is taken to be a realization of a random process, often taken to be a Gaussian process with a Gaussian correlation function. One can imagine making predictions for this model by generating many random realizations of the potential, simulating the dynamics of the inflaton fields, and using the result to make predictions for observable quantities (scalar and tensor power spectra, non-Gaussianity, etc.) In a high-dimensional space, simulating a realization throughout an entire box is computationally intractable. Fortunately, we need the simulation only along the inflaton path. 

The most well-developed method for making predictions in such models is built on methods of random matrix theory (RMT) \citep[e.g.,][and references therein]{aazami,marsh,dias1,dias,masaki,paban,feng}. These methods are based on approximations and assumptions that may be problematic \citep{freivogel,wang,easther,masoumi}. Among other issues, these methods can lead to solutions that do not have continuous derivatives, which can lead to problems in making predictions of non-Gaussianity \citep{bjorkmo}, although adaptations are possible \citep{battefeld}.

Other methods besides RMT have been used to explore multifield inflation in random potentials. Some involve simulating the potential in the vicinity of a single point \citep{masoumi,bjorkmo}. Other methods involve random walks \citep{tye}, Bayesian networks \citep{price}, and machine learning \citep{rudelius}.

In this paper, we present a method of simulating trajectories in a random landscape directly, without recourse to the approximations involved in techniques such as RMT. At each step along the numerical solution of the inflaton equations of motion, we simulate the potential, taking into account the information provided by the simulations at previous steps along the path. Similar methods are found in \cite{blanco-pillado}.  

At each step in our simulation, the values of $V$ and its gradient are generated at the next point along the inflaton trajectory. These values are drawn from the conditional probability distribution that takes account of both the correlation function of the Gaussian random process from which the full function $V$ is assumed to be drawn and all of the previous potential and gradient values that have been sampled. The result is  equivalent to simulating the full potential function in advance and numerically solving for the trajectory but is much less computationally expensive.

In order to keep the computation from becoming intractable, we periodically ``forget'' information that no longer significantly influences the simulation. In principle, this forgetting step breaks the precise equivalence between our method and simulating the entire potential in advance, but we are able to quantify the degree of approximation involved and show that it is small.

We illustrate our method by computing many  simulated trajectories and using the results to predict distributions of various observable quantities, including the adiabatic, isocurvature, and tensor power spectra. In these illustrations, we consider relatively low values of $N$ and focus on a particular portion of parameter space. With further code optimization and/or the use of more computation time, we expect to extend this work to larger numbers of dimensions and a wider range of input parameters.

A central feature of our approach is that we do not restrict our attention in advance to certain categories of trajectory (e.g., slow-roll slow turn or rapid turn). Rather, we simulate trajectories drawn at random from the probability distribution specified by the Gaussian random process for $V$, and let this determine the frequency with which these different types of trajectory appear. The one exception is that we impose \textit{a priori} the  constraint that inflation ends, by only considering trajectories that end at a minimum at which $V=0$. This constraint is justified on anthropic grounds  \citep[e.g.,][]{barrow,efstathiou,susskind}: in the landscape models under consideration, most of the universe inflates forever or recollapses rapidly, but we live in one of the rare patches that did neither.

The paper is structured as follows: requisite multi-field inflationary theory is presented in the remainder of Section \ref{sec:introduction}, the generation of Gaussian random potentials is formalized in Section \ref{sec:formalism}, our original code and data compression algorithms are detailed in Section \ref{sec:computational_process}, results are given in Section \ref{sec:results}, and we conclude with Section \ref{sec:conclusions}.

\subsection{Multi-field Inflation} \label{sec:multifield_inflation}

As customary in inflationary theory, we suppose there existed some ``inflaton'' field $\boldsymbol{\phi}(t, \boldsymbol{x})$ which permeated the early universe with time evolution governed by its associated potential function $V(\boldsymbol{\phi})$. We assume that the inflaton field was spatially homogeneous with small perturbations, i.e., $\boldsymbol\phi(t, \boldsymbol{x}) \approx \boldsymbol\phi(t) + \delta\boldsymbol\phi(t,\boldsymbol{x})$  In general, $\boldsymbol{\phi} = (\phi^{(0)}, \phi^{(1)}, ... , \phi^{(N-1)})$ is a vector of $N$ components while $V$ is scalar-valued. In addition to its existence, we assert that the inflaton potential dominates the energy density of the early universe so that its characteristics may determine the cosmic expansion history at early times. The Klein-Gordon equation for the evolution of $\boldsymbol{\phi}$ is
\beq \label{eq:inflaton}
\ddot{\boldsymbol\phi}+3H\dot{\boldsymbol\phi}+\nabla_{\phi} V=0,
\eeq
where overdots signify derivatives with respect to cosmic time $t$, $H$ is the Hubble parameter, and Planck units $c=\hbar=8\pi G=1$ are implemented throughout. We assume a flat field-space metric \citep[e.g.,][]{langlois}; generalization of our method to any given metric should be possible.

Equation \eqref{eq:inflaton} is written componentwise as
\beq \label{eq:inflaton_cw}
\ddot{\phi}^{(\alpha)} + 3H\dot{\phi}^{(\alpha)} + \frac{\partial V}{\partial\phi^{(\alpha)}} = 0,
\eeq
for $\alpha=0,\ldots,N-1$.
As the well-known analogy goes, this system of equations is equivalent to a ball rolling in the potential $V$ with a friction force given by the Hubble parameter.

The Hubble parameter is related to the field by the Friedmann equation,
\beq \label{eq:friedmann}
3H^2 = \frac{1}{2}\big|\dot{\boldsymbol\phi} \big|^2 + V.
\eeq

Given the potential, the second-order differential equation \eqref{eq:inflaton_cw} can be solved numerically with initial conditions $\boldsymbol\phi_i = \boldsymbol\phi(t_i)$ and $\dot{\boldsymbol\phi}_i = \dot{\boldsymbol\phi}(t_i)$. Often, the initial rate of change is set to slow-roll equilibrium so that $\ddot{\boldsymbol\phi}_i=0$. By \eqref{eq:inflaton}, the approximation is
\beq \label{eq:slow_roll_1}
\dot{\boldsymbol\phi}_i \approx -\frac{\nabla_{\phi}V_i}{3H_i}.
\eeq
Furthermore, if $|\dot{\boldsymbol\phi}|^2$ is assumed to be negligibly small, then \eqref{eq:friedmann} reduces to $3H^2 \approx V$ and the slow-roll approximation is
\beq \label{eq:slow_roll_2}
\frac{\dot{\boldsymbol\phi_i}}{H_i} \approx -\frac{\nabla_{\phi}V_i}{V_i}.
\eeq
For numerical and analytic purposes, it is often useful to measure time in terms of the expansion of the universe \citep[e.g.,][]{mmc}.
During inflation, the scale factor $a(t)$ grows approximately exponentially, so a natural unit of time is the ``number of $e$-folds''
\beq \label{eq:N_e}
N_e(t) = \ln{\left(\frac{a(t)}{a(t_i)}\right)},
\eeq
where $t_i$ is some arbitrary start time of inflation. Since $H=\dot{a}/a$, the left-hand side of \eqref{eq:slow_roll_2} then simplifies to $d\boldsymbol\phi_i/dN_e$ and, under slow roll, some algebra transforms \eqref{eq:inflaton_cw} into
\beq \label{eq:inflaton_cw_adv}
\frac{d^2\phi^{(\alpha)}}{dN_e^2} + (3-\epsilon)\frac{d\phi^{(\alpha)}}{dN_e} + \frac{1}{H^2}\frac{\partial V}{\partial\phi^{(\alpha)}} = 0,
\eeq
in which we have defined the slow-roll parameter $\epsilon$ as
\beq \label{eq:epsilon}
\epsilon \equiv -\frac{\dot{H}}{H^2} = \frac{1}{2}\left|\frac{d\boldsymbol\phi}{dN_e}\right|^2,
\eeq
signaling the end of inflation at $t_e$ when $\epsilon\uparrow 1$. Note that $H$ can be computed at any time step by 
\beq \label{eq:H}
H^2 = \frac{V}{3-\epsilon}.
\eeq
Once numerical integration finishes at $\epsilon=1$, the values of $\boldsymbol\phi$, $\dot{\boldsymbol\phi}$, $V$, $\nabla_{\phi}V$, $\epsilon$, and $H$ are known at discrete times between $t_i$ and $t_e$. We refer to this set of information as the inflaton field \textit{trajectory}.

\subsection{Power Spectra} \label{sec:power_spectra}

Various observable quantities can be extracted from the trajectory, such as the power spectra of adiabatic curvature (\change{scalar}) fluctuations $\mathscr{P}_S$, tensor fluctuations $\mathscr{P}_T$, isocurvature fluctuations $\mathscr{P}_{\rm{iso}}$ and the non-Gaussianity of the CMB \citep{price}. In this paper, we will focus on the power spectra.

Following \cite{mmc}, \change{curvature} perturbations evolve with the mode matrix $\boldsymbol\psi$, an expansion of the comoving fluctuations $\delta\boldsymbol\phi$ in the inflaton field. For any given mode $k$, the mode matrix obeys
\beq \label{eq:mode_matrix}
\frac{d^2\boldsymbol\psi}{dN_e^2} + (1-\epsilon)\frac{d\boldsymbol\psi}{dN_e} + \bigg( \frac{k^2}{a^2 H^2} - 2 + \epsilon \bigg)\boldsymbol\psi + \boldsymbol{C}\boldsymbol{\psi} = 0,
\eeq
where the coupling matrix $\boldsymbol{C}$ is defined as
\begin{multline} \label{eq:coupling_matrix}
C^{(\alpha\beta)} = \frac{1}{H^2}\bigg( \frac{d\phi^{(\alpha)}}{dN_e}\frac{\partial V}{\partial \phi^{(\beta)}} + \frac{d\phi^{(\beta)}}{dN_e}\frac{\partial V}{\partial \phi^{(\alpha)}} + \frac{\partial^2 V}{\partial\phi^{(\alpha)}\partial\phi^{(\beta)}}\bigg)\\
+ (3-\epsilon)\frac{d\phi^{(\alpha)}}{dN_e}\frac{d\phi^{(\beta)}}{dN_e}.
\end{multline}
Given conditions on $\boldsymbol\psi$ and $d\boldsymbol\psi/dN_e$ at $t_i$, equation \eqref{eq:mode_matrix} can be solved numerically up through $t_e$. At any given time, $\mathscr{P}_S(k)$ is computed as
\beq \label{eq:P_S}
\mathscr{P}_S(k) = \frac{k^3}{4\pi^2\epsilon}\frac{1}{a^2}\boldsymbol\Psi\boldsymbol\Psi^*,
\eeq
where $\boldsymbol\Psi = \dot{\boldsymbol\phi}\cdot\boldsymbol\psi/|\dot{\boldsymbol\phi}|$.
and $*$ is the conjugate transpose. To be clear, the power spectrum, as with the mode matrix, is a function of both wavenumber and time. We are interested in the shape of $\mathscr{P}_S(k)$ at the end of inflation, as this spectrum \change{can be used to compare with observations.\footnote{\change{For single-field inflation, the power spectrum at any given $k$ can be evaluated at the moment that wavenumber leaves the horizon, as superhorizon evolution is trivial, but for the multifield models considered here nontrivial evolution can occur after this time. The tensor spectrum, on the other hand, can be computed at horizon-crossing even for these models.}}
We assume standard, rapid reheating at the end of inflation.}

We expect the power spectrum to resemble a power law of amplitude $A_S$ and index $n_S$ around the pivot scale $k_0 = 0.05\ \text{Mpc}^{-1}$:
\beq \label{eq:P_S_power_law}
\mathscr{P}_S(k)\bigg|_{t_e} \approx A_S\bigg( \frac{k}{k_0} \bigg)^{n_S-1}.
\eeq
For any reasonable inflationary model, $\mathscr{P}_S$ can be evolved through time at several modes evenly log-spaced around $k_0$ via \eqref{eq:mode_matrix}, then fit to \eqref{eq:P_S_power_law} at $t_e$. The best fit parameters $A_S$ and $n_S$ can then be used to gauge the viability and predictive power of the choice of inflaton potential and initial conditions.

Similarly, the power spectrum of isocurvature perturbations can be computed at any point in time as
\beq \label{eq:P_iso}
\mathscr{P}_{\mathrm{iso}}(k) = \frac{k^3}{4\pi^2\epsilon}\frac{1}{a^2}\sum_{\alpha}^{N-1}\boldsymbol\Psi_{\alpha}\boldsymbol\Psi_{\alpha}^*
\eeq
where $\boldsymbol\Psi_{\alpha} = \hat{\boldsymbol{s}}_{\alpha}\cdot\boldsymbol\psi$ for each unit vector $\hat{\boldsymbol{s}}_{\alpha}$ orthogonal to $\dot{\boldsymbol\phi}$. As with the \change{curvature} power spectrum, we sample $\mathscr{P}_{\rm{iso}}$ around $k_0$, then fit to a power law,
\beq \label{eq:P_iso_power_law}
\mathscr{P}_{\mathrm{iso}}(k)\bigg|_{t_e} \approx A_{\mathrm{iso}}\bigg( \frac{k}{k_0} \bigg)^{n_{\mathrm{iso}}-1}.
\eeq
The isocurvature spectrum amplitude is typically expressed as the isocurvature-to-scalar ratio $\beta_{\rm{iso}}(k_0) \equiv A_{\rm{iso}}/(A_{\rm{iso}} + A_S)$, so we will report extract and report this quantity from our own model.

Following \cite{langlois}, the power spectrum of tensor perturbations is easier to compute:
\beq \label{eq:P_T}
\mathscr{P}_T(k) = \frac{2}{\pi^2}H(t(k))^2.
\eeq
The one-to-one and onto function $t(k)$ yields the time at which a scale of magnitude $k$ last exited the horizon, most easily expressed as its inverse, $k(t) = a(t)H(t)$. Thus, in this definition, $\mathscr{P}_T(k)$ gives by default the power at the end of inflation. As with the \change{curvature} spectrum, the tensor spectrum can be sampled at several wavenumbers around $k_0$ and fit to another anticipated power law,
\beq \label{eq:P_T_power_law}
\mathscr{P}_T(k) \approx A_T\bigg( \frac{k}{k_0} \bigg)^{n_T}.
\eeq
The tensor perturbation amplitude is often expressed in terms of the tensor-to-scalar ratio $r$, which is defined as the power law ratio for any given $k$,
\beq \label{eq:r}
r(k) = \frac{\mathscr{P}_T(k)}{\mathscr{P}_S(k)} = \frac{A_T}{A_S}\bigg( \frac{k}{k_0} \bigg)^{n_T-n_S+1}.
\eeq
We will quote tensor-to-scalar ratios at the scale $k_t = 0.002\ \text{Mpc}^{-1}$.

Constraints on $A_S$, $n_S$, $r(k_t)$, and $\beta_{\rm{iso}}(k_0)$ have been found by recent Planck missions and are summarized in Table \ref{table:planck_constraints} \citep{planck}.

\change{When computing all of the power spectra described above, we follow the procedures used in the MultiModeCode software \citep{mmc}, and we have confirmed that the two methods give matching results in particular cases such as a quadratic potential.}

\begin{table}
\begin{tabular}{|p{4.0cm}|p{4.0cm}|} 
   \hline
   \textbf{Spectral Quantity} & Constraint (Planck 2018)\\
   \hline
   \hline
   $\log(10^{10}A_S)$  &  $3.040 \pm 0.016$\\
   \hline
   $n_S$  &  $0.9626 \pm 0.0057$\\
   \hline
   $r(k_t)$ & $< 0.1$\\
   \hline
   $\beta_{\rm{iso}}(k_0)$ & $< 0.47$\\
   \hline
\end{tabular}
\caption{Planck constraints on \change{curvature}, tensor, and isocurvature fluctuations \citep{planck}.}
\label{table:planck_constraints}
\end{table}

\section{Formalism} \label{sec:formalism}

The review in Section \ref{sec:introduction} was independent of the shape of inflaton potential, but a definition of $V(\boldsymbol\phi)$ is the heart of every inflationary model. In this paper, we assume that $V(\boldsymbol\phi)$ is a realization of an $N$-dimensional Gaussian random field with mean zero, coherence scale $s$, and energy scale $V_{\star}$ \citep{tye}. These three parameters ($s,V_{\star},N$) do not uniquely define a potential, but rather characterize the statistical properties of a family of random potentials. The coherence scale is a characteristic distance in $\boldsymbol\phi$-space below which values of $V$ are strongly correlated. 
The inflationary energy scale $V_{\star}$ is the standard deviation of $V$. In order to produce scalar fluctuations at levels that approximately match observations, $V_{\star}$ should be of order $10^{-5}\ E_{\text{Pl}}^{4}$. $N$ is the dimension of $V$, or the number of components of $\boldsymbol{\phi}$.

\subsection{Generating Correlated Points} \label{sec:generating_correlated_points}

\begin{figure*} 
    \centerline{
    \includegraphics[width=7.0in]{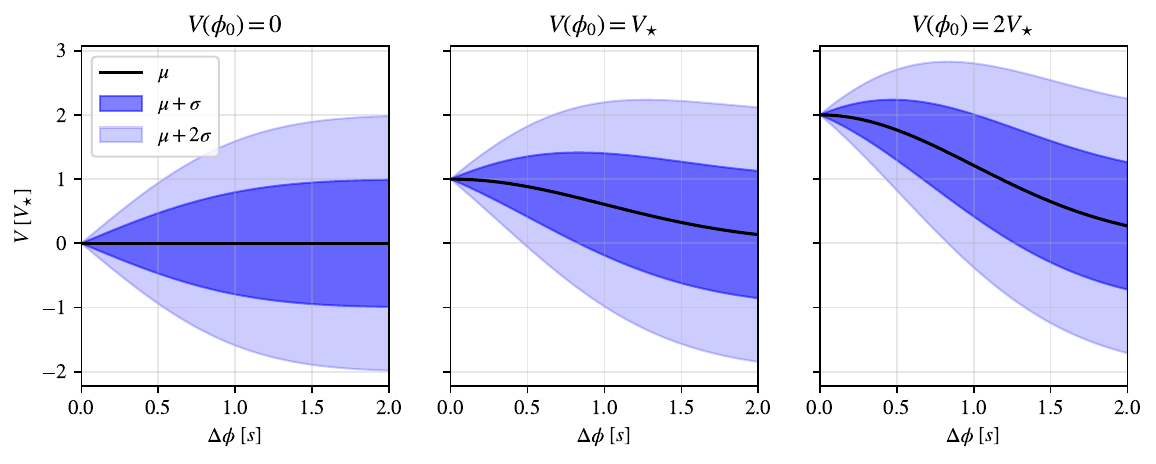}
    }
    \caption{Distribution of possible $V(\boldsymbol{\phi}_1)$ values as a function of $\Delta\phi$, following the in-text example. Given $V(\boldsymbol{\phi}_0) = aV_{\star}$ and seeking to generate $V(\boldsymbol{\phi}_1)$ at some distance $\Delta\phi = |\boldsymbol{\phi}_0 - \boldsymbol{\phi}_1|$ away, the figure shows the mean $\mu$ (black line) and $1\sigma$, $2\sigma$ intervals of the distribution from which $V(\boldsymbol{\phi}_1)$ is drawn for a range of separations. When $\Delta\phi$ is small, the points are close together and $V_1$ will be highly constrained around $V_0$. However, the influence of $V_0$ on $V_1$ decays as $\Delta\phi$ grows.}
    \label{fig:mu_plus_skew_visual_aid}
\end{figure*}

In order to simulate a trajectory, one could begin by simulating the potential $V$ throughout an $N$-dimensional box of $\boldsymbol{\phi}$-space in which the trajectory is expected to lie, but for large $N$ this is extremely inefficient. It is far more efficient to generate $V$ only along the trajectory; potential values elsewhere are irrelevant and unnecessary for generating new values.
 We describe and implement this method of simulating $V$ along the path of $\boldsymbol\phi$ in the following sections.

Generating points on a Gaussian random field is a constrained random process; every new point is generated randomly \textit{constrained by} every previously generated point. This is necessary to ensure that the potential has the desired correlation structure -- in particular, that it is appropriately smooth on scales less than $s$.

Suppose that $V$ is known at $P$ individual ``past'' points in $\boldsymbol{\phi}$-space, and that those known values are grouped into a vector 
\begin{equation*}
    \boldsymbol{v}_{\mathcal{P}}= (V(\boldsymbol\phi_0), V(\boldsymbol\phi_1), \hdots, V(\boldsymbol\phi_{P-1})).
\end{equation*}
Now, we want to generate new potential values at $F$ ``future'' points,  
\begin{equation*}
    \boldsymbol{v}_{\mathcal{F}} = (V(\boldsymbol\phi_{P}), V(\boldsymbol\phi_{P+1}), \hdots, V(\boldsymbol\phi_{P+F-1})),
\end{equation*}
constrained by the information in $\boldsymbol{v}_{\mathcal{P}}$. First, we need to define the \textit{covariance matrix} $\boldsymbol\Gamma$ of constraints on $V$ as an $(P+F) \times (P+F)$ matrix with elements
\begin{align} 
    \Gamma_{ij} &= \langle V(\boldsymbol\phi_i)V(\boldsymbol\phi_j) \rangle \equiv \langle V_i V_j \rangle \\
    &= V_{\star}^2 \exp \left( -\frac{|\boldsymbol{\phi}_i - \boldsymbol{\phi}_j|^2}{2s^2} \right) \label{eq:covariance_matrix}
\end{align}  
where $i,j$ index $\boldsymbol{v}=(\boldsymbol{v}_{\mathcal{P}},\boldsymbol{v}_{\mathcal{F}})$. This symmetric, positive-definite matrix is subdivided into blocks corresponding to the type of information stored:
\begin{equation}
    \boldsymbol\Gamma =
    \begin{bmatrix}
    \boldsymbol\Gamma_{\mathcal{PP}} & \boldsymbol\Gamma_{\mathcal{PF}}\\
    \boldsymbol\Gamma_{\mathcal{FP}} & \boldsymbol\Gamma_{\mathcal{FF}}
    \end{bmatrix}.
    \label{eq:Gamma_blocks}
\end{equation}
The upper left block $\boldsymbol{\Gamma}_{\mathcal{PP}}$ contains covariances of past points, the offdiagonal blocks $\boldsymbol{\Gamma}_{\mathcal{PF}}=\boldsymbol{\Gamma}_{\mathcal{FP}}^T$ contain cross-covariances between past and future points, and $\boldsymbol{\Gamma}_{\mathcal{FF}}$ contains covariances of future points \citep{numericalrecipes}. The vector $\boldsymbol{v}_\mathcal{F}$ is drawn from a Gaussian distribution \citep{hoffman,rybicki} with mean
\beq \label{eq:mu}
\boldsymbol\mu = \boldsymbol\Gamma_{\mathcal{FP}}\boldsymbol\Gamma_{\mathcal{PP}}^{-1}\boldsymbol{v}_{\mathcal{P}}
\eeq
and covariance matrix
\beq \label{eq:conditional_covariance_matrix}
\boldsymbol\Gamma_{\mathcal{C}} = \boldsymbol\Gamma_{\mathcal{FF}} - \boldsymbol\Gamma_{\mathcal{FP}}\boldsymbol\Gamma_{\mathcal{PP}}^{-1}\boldsymbol\Gamma_{\mathcal{PF}}.
\eeq
Explicitly,
\beq \label{eq:v_F}
\boldsymbol{v}_{\mathcal{F}} = \boldsymbol\mu + \boldsymbol{L}_{\mathcal{C}}\boldsymbol{y}.
\eeq
where $\boldsymbol\Gamma_{\mathcal{C}} = \boldsymbol{L}_{\mathcal{C}}\boldsymbol{L}_{\mathcal{C}}^T$ is the lower-triangular Cholesky decomposition and $\boldsymbol{y}$ is a vector of $F$ independent standard normal values drawn from 
$\mathcal{N}(\boldsymbol0,\boldsymbol{1})$.

As an illustrative example, suppose we sample the potential at $\boldsymbol\phi_0$ to be $V(\boldsymbol\phi_0) = a V_{\star}$, and \change{we}  seek to generate a second point, $V(\boldsymbol\phi_1)$, constrained by $V(\boldsymbol\phi_0)$. If the points are separated in $\boldsymbol{\phi}$-space by $|\boldsymbol\phi_0 - \boldsymbol\phi_1| = \Delta \phi$, the covariance matrix is
\beq \label{eq:covariance_matrix_ex}
\boldsymbol\Gamma = 
\begin{bmatrix}
    V_{\star}^2 & V_{\star}^2 \exp( -\Delta\phi^2/2s^2 )\\
    V_{\star}^2 \exp( -\Delta\phi^2/2s^2 ) & V_{\star}^2
\end{bmatrix}.
\eeq
Since the number of future points is $F=1$, the mean is a scalar,
\beq \label{eq:mu_ex}
\mu = a V_{\star} \exp\bigg( -\frac{\Delta\phi^2}{2s^2} \bigg),
\eeq
as is the conditional covariance matrix,
\beq \label{eq:Gamma_C_ex}
\Gamma_{\mathcal{C}} = V_{\star}^2\bigg[ 1 - \exp\bigg(-\frac{\Delta\phi^2}{s^2}\bigg) \bigg].
\eeq
The new point is then generated as $V(\boldsymbol{\phi}_1) = \mu + y\sqrt{\Gamma_{\mathcal{C}}}$, where $y$ is drawn from a standard normal distribution. Figure \ref{fig:mu_plus_skew_visual_aid} plots $\mu$ as a function of $\Delta\phi$ with one and two-sigma regions for $a = 0, 1$, and $2$. As $\Delta\phi$ grows, $\mu$ trends from $V(\boldsymbol\phi_0)$ toward zero, the unconstrained mean, and the standard deviation increases from zero toward $V_{\star}$. In other words, the value of $V_1$ is highly constrained around $V_0$ when $\boldsymbol\phi_1$ is close to $\boldsymbol\phi_0$, but the constraint weaken as the separation grows.

We also store derivatives of the potential and their covariances with other information. For example, if one element $v_i$ of $\boldsymbol{v}$ is a derivative with respect to $\phi^{(\alpha)}$, $\partial V/\partial\phi^{(\alpha)}$, and another element $v_j$ is a value of $V$, then
\begin{align}
\Gamma_{ij} &= \bigg\langle \frac{\partial V_i}{\partial\phi_i^{(\alpha)}} V_j \bigg\rangle = \frac{\partial}{\partial\phi_i^{(\alpha)}} \langle V_i V_j \rangle\\
&= \Bigg( \frac{\phi_j^{(\alpha)} - \phi_i^{(\alpha)}}{s^2} \Bigg) \langle V_i V_j \rangle \label{eq:derivative_correlation_ex}
\end{align}
All correlations between first and second derivatives of $V$ can be derived in a similar way, albeit with tedious algebra.

We emphasize that generating values of $V$ and/or its derivatives along a trajectory by keeping track of the full conditional probability distribution at each point is precisely equivalent to generating an entire sample function $V$ in advance and then sampling it at the given points. The reason is simply that the joint probability distribution of a set of $n$ random variables is the same as the product of the appropriate conditional probabilities: $p(x_1,x_2,\ldots, x_n) = p(x_1)p(x_2|x_1)p(x_3|x_1,x_2)\ldots$. If we wanted to, we could go back at any later time and sample other quantities from the appropriate conditional probability distribution. For example, if we have simulated $V$ and $\nabla V$ at a series of points, we could later go back and simulate the Hessian matrix $\{\partial^2 V/(\partial \phi^{(\alpha)}\partial \phi^{(\beta)})\}$ at any given point by drawing from the appropriate conditional probability distribution, and we would find that it had all the expected properties -- e.g., the matrix would be symmetric, because the random variables $\partial^2 V/(\partial \phi^{(\alpha)}\partial \phi^{(\beta)})$ and $\partial^2 V/(\partial \phi^{(\beta)}\partial \phi^{(\alpha)})$ are 100\% correlated.

\subsection{The Minimum Condition} \label{sec:minimum_condition}

\begin{figure*} 
    \centerline{
    \includegraphics[width=7.0in]{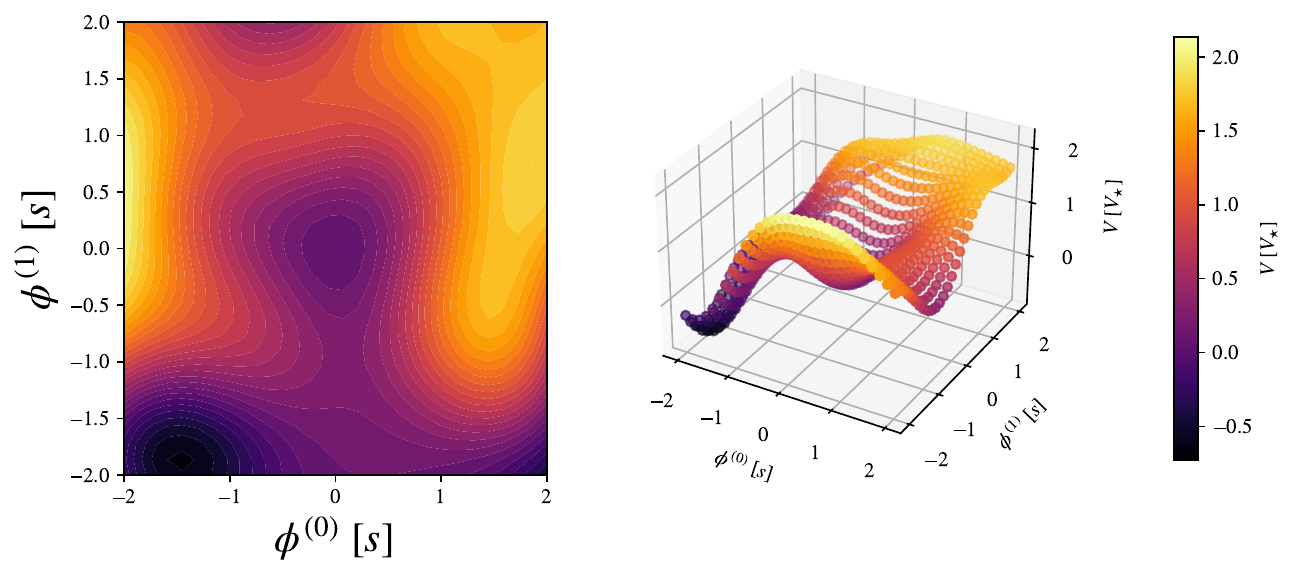}
    }
    \caption{A $2$-dimensional Gaussian random potential $V$ generated at nodes on a $40 \times 40$ grid evenly spaced in $\phi^{(0)}, \phi^{(1)}$. \textit{Left}: a ``birds-eye view'' of the potential isocontours, colored by potential value. \textit{Right}: the same potential function as a traditional 3-D projection, with $V$ as the height. In both panels, the imposed minimum conditions are clearly visible. Note that we do not employ these techniques of generating potential values throughout an entire rectangular domain when solving the trajectory, but do so here for illustrative purposes.}
    \label{fig:example_minimum}
\end{figure*}

Generic inflaton trajectories do not lead to universes like our own: the solution settles into a minimum of $V$ that is generically quite far from zero, leading to a universe that either inflates forever or rapidly recollapses. When comparing with observation, we wish to look at models that are anthropically constrained \citep[e.g.,][]{barrow,efstathiou,susskind} to end in a minimum whose value is extremely close to zero.

We implment this anthropic constraint by imposing the following $N^2+N+1$ conditions on $V$ before the trajectory is solved:
\begin{align}
    V(\boldsymbol0) &= 0 \label{min_condition_1}\\
    \frac{\partial V}{\partial\phi^{(\alpha)}}(\boldsymbol0) &= 0 \text{ for all }\alpha \label{min_condition_2}\\
    \frac{\partial^2 V}{\partial (\phi^{(\alpha)})^2}(\boldsymbol0) &> 0 \text{ for all }\alpha \label{min_condition_3}\\
    \frac{\partial^2 V}{\partial\phi^{(\alpha)}\partial\phi^{(\beta)}}(\boldsymbol0) &= 0 \text{ for all }\alpha \ne \beta \label{min_condition_4}
\end{align}
These conditions ensure that the potential is zero, flat, and concave up at the origin. More precisely, we want the Hessian $\boldsymbol{H}$ to be positive-definite at the origin. Since there are no preferable locations in a Gaussian random field, we could have chosen any location for the minimum, but chose the origin for convenience. Constraints \eqref{min_condition_3} and \eqref{min_condition_4} are equivalent to diagonalizing $\boldsymbol{H}$ and forcing its diagonal elements to be positive. Physically, we reorient our axes to align with the principal directions of curvature. 

Figure \ref{fig:example_minimum} shows an example Gaussian random potential subject to the minimum conditions at the origin. Smoothly connected valleys and peaks are visible in the landscape. Notice that there are some locations from which $\boldsymbol{\phi}$ would \textit{not} evolve toward the origin (e.g. the southwest and north, on the left panel), instead meandering away to some adjacent minimum. We discard these runs as unphysical universe on anthropic grounds since $V_e \ne 0$. However, the likelihood that $\boldsymbol{\phi}$ converges to the minimum at the origin increases as $|\boldsymbol{\phi}_i|$ decreases; in this case, $|\boldsymbol\phi_i|<s$ almost guarantees convergence.

Note that at $V = 0$, our simulated minima are likely to be shallower than more-common minima at $V < 0$. It is likely, especially with more inflaton components, that at the concavity is only slightly greater than zero along at least one principal direction. For these common potentials, the minimum appears more like a saddle point, and $\boldsymbol\phi$ ``momentum'' can easily carry the inflaton field \textit{through} the shallow minimum and out the other side (also leading to a discarded run).

\section{Computational Process} \label{sec:computational_process}

We developed the \texttt{bushwhack} code to solve the equations of motion for the inflaton field in an $N$-dimensional Gaussian random potential. Careful measures were implemented to simplify and compress data related to the random generation of $V$, the most computationally intensive feature.

\subsection{The Program} \label{sec:program}

\begin{figure*} 
    \centerline{\includegraphics[width=7.0in]{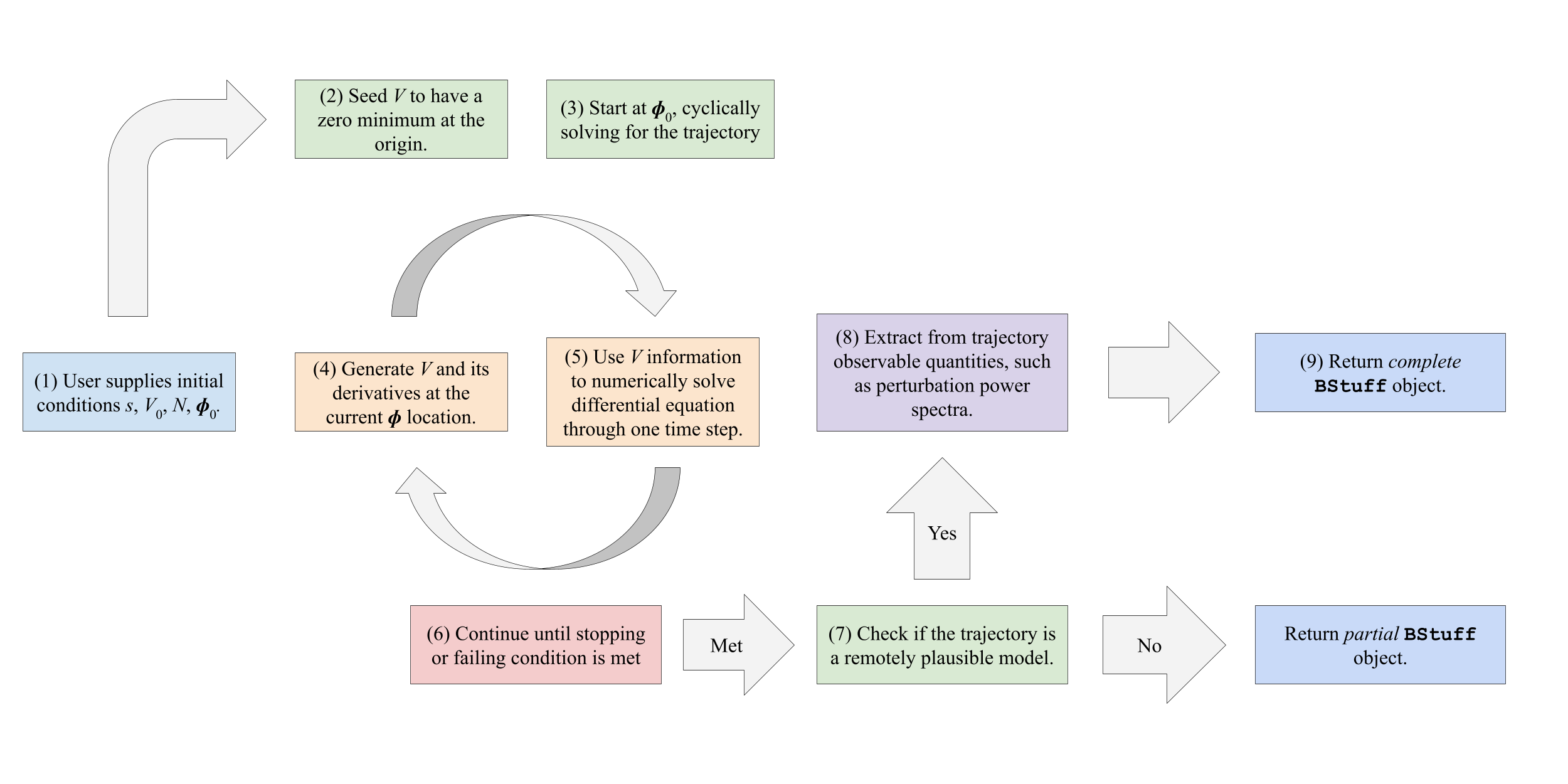}}
    \caption{Flowchart for \texttt{bushwhack} code illustrating the chronology of computational processes involved in solving for the trajectory of $\boldsymbol{\phi}$ and extracting observable quantities. The user supplies initial conditions, a minimum is seeded at the origin (an anthropic constraint), the inflaton field is evolved until a stopping criterion triggers, and \change{curvature}, tensor, and isocurvature spectra are extracted.}
    \label{fig:bushwhack_flowchart}
\end{figure*}

Figure \ref{fig:bushwhack_flowchart} shows the general progression of the \texttt{bushwhack} program, from the input of initial conditions to the output information about predicted observable quantities. We detail each step below.

\textit{Supply initial conditions.} First, we characterize the potential by choosing its parameters, $s$, $V_{\star}$, and $N$. The initial $\phi^{(\alpha)}$ values are determined by randomly selecting a location near the origin at some user-inputted radius $\phi_i \equiv |\boldsymbol{\phi}_i|$. This random choice of initial $\boldsymbol\phi$ is necessary to collect a representative sample of trajectories. Given that we diagonalize the Hessian matrix at the origin (see Section \ref{sec:minimum_condition}), forcing $\boldsymbol\phi_i = (\phi_i,0,0,\hdots,0)$, for instance, would yield paths which always start near a principal axis of curvature. The inflaton ``velocity'' $\dot{\boldsymbol\phi}$ need not be user-supplied, since its value will be automatically computed from $V_i$ and $\nabla_{\boldsymbol\phi}V_i$ via equation \eqref{eq:slow_roll_2}.

\textit{Seed the minimum.} Before solving for a trajectory, the code enforces conditions \eqref{min_condition_1}-\eqref{min_condition_4} to allow for the possibility that the solution is physically plausible. In particular, condition \eqref{min_condition_3} is imposed by randomly simulating $N$ independent second derivative values, taking their absolute values to force them positive, and storing them as the diagonals of the Hessian of $V$ at the origin. At this step, many arrays are initialized such as $\boldsymbol\Gamma$, $\boldsymbol{v}$, $\boldsymbol\phi$, and $d\boldsymbol\phi/dN_e$. The imposed constraints at the origin will remain the first $N^2+N+1$ elements of $\boldsymbol{v}$ throughout the simulation and constrain the rest of the potential via a corresponding upper-left block of $\boldsymbol\Gamma=\langle \boldsymbol{v}\boldsymbol{v}^T \rangle$.

\textit{Simulate--solve cycle.} Beginning at $\boldsymbol\phi_i$, $N+1$ data points are randomly generated, $V_i$ and $\nabla_{\phi}V_i$. Those values are then substituted in equation \eqref{eq:inflaton_cw_adv} and the field is solved numerically through one time step to $\boldsymbol\phi_{i+1}$ via a fourth-order Runge-Kutta solution algorithm. Then $V_{i+1}$ and $\nabla_{\phi}V_{i+1}$ are generated and so continues the ``simulate--solve cycle''. At any given time step, the potential is generated exactly along the path evolution of $\boldsymbol\phi$, constrained by all previously generated data stored in $\boldsymbol\Gamma$. In the language of Section \ref{sec:generating_correlated_points}, past constraints define $\boldsymbol\Gamma_{\mathcal{PP}}$, which gets larger and more burdensome as the simulation goes on, while $\boldsymbol\Gamma_{\mathcal{FF}}$ remains an $(N+1)\times(N+1)$ square representing the new potential and gradient values at the current position.

The most computationally expensive steps in this cycle are those involving $\boldsymbol{\Gamma}_\mathcal{PP}^{-1}$ (Equations \ref{eq:mu} and \ref{eq:conditional_covariance_matrix}). Because this matrix is symmetric positive definite, these are done via Cholesky decomposition and solving systems of equations with a triangular matrix, which is much faster and numerically stable. This decomposition can be updated incrementally to account for the new rows and columns added in each time step; it does not have to be done from scratch.

It is also worth noting that this matrix has a large \textit{condition number}: many of the constraints are nearly redundant, leading to nearly null eigenvectors. To ensure numerical stability, one can add a small multiple of the identity to the matrix with negligible impact on subsequent potential generation.

\begin{figure*} 
    \centerline{
    \includegraphics[width=7.5in]{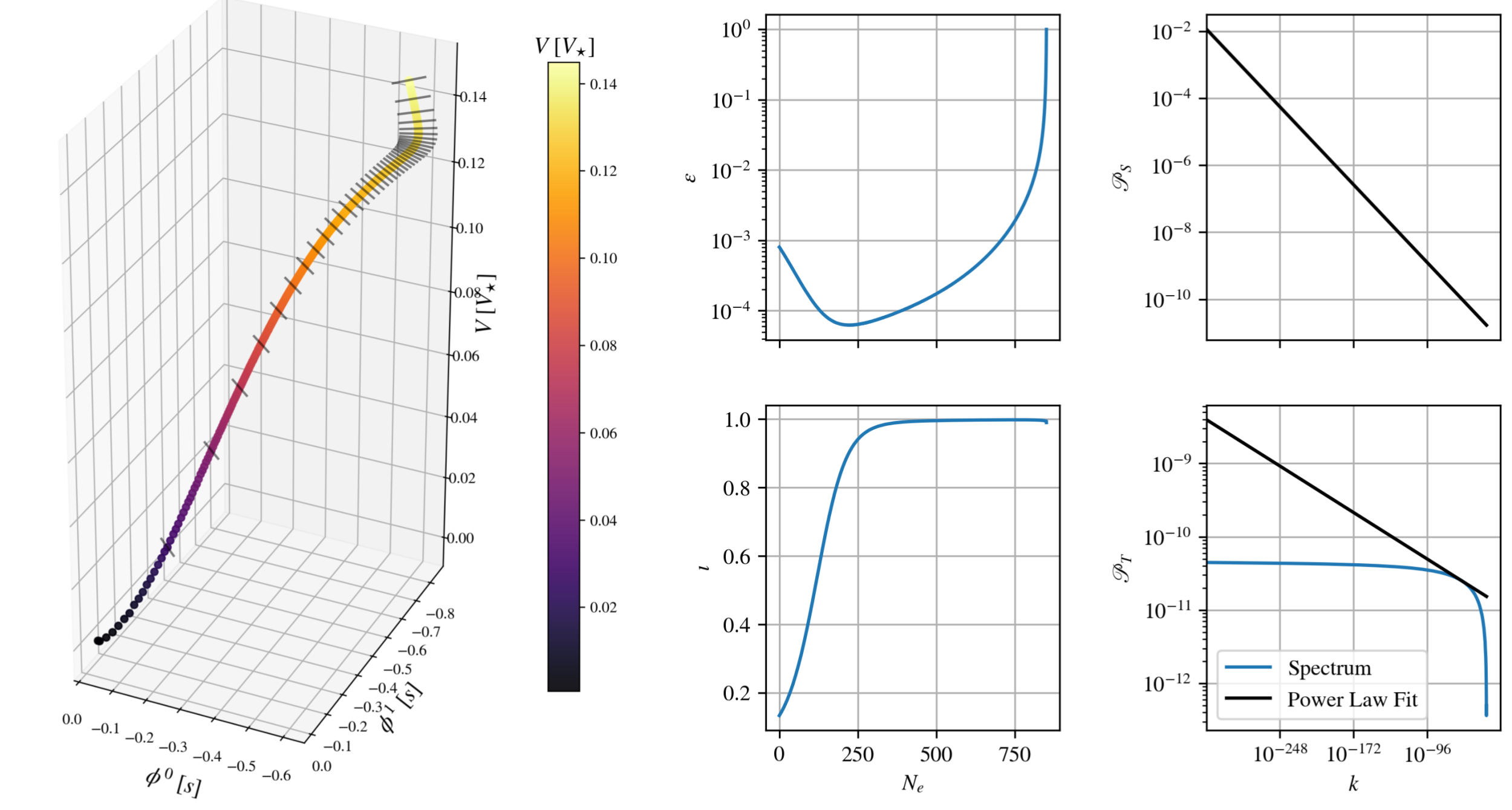}
    }
    \caption{A sample trajectory in a Gaussian random potential with parameters $s=30$, $V_{\star}^2=5\times10^{-9},$ and $N=2$. \textit{Left}: the inflaton field meanders in the direction of $\nabla_{\phi}V$ (hatches along the curve) from some initial $\phi_i=s$ to the seeded minimum at the origin. \textit{Middle}: slow roll parameter $\varepsilon$ and inwardness $\iota$, tracked to determine when to stop evolving $\boldsymbol{\phi}$. In this case, $\epsilon$ triggers when it accelerates to $1$, a sign that the trajectory converged to the origin. The inwardness parameter detects that $\boldsymbol{\phi}$ moved nearly perpendicular to the origin at first, but never triggers. \textit{Right}: power-law fits to \change{curvature} (top) and tensor (bottom) power spectra. Analytic curve for $\mathscr{P}_T(k)$ is shown in blue. Note that the approximation is only valid near $k_0=0.05~\text{Mpc}^{-1} \approx 1.31\times 10^{-58}\ l_{\rm{Pl}}^{-1}$.}
    \label{fig:example_trajectory}
\end{figure*}

\textit{Stopping conditions.} Multiple stopping conditions are monitored at every time step to supervise the evolution of $\boldsymbol\phi$. The first slow-roll parameter, $\epsilon$, turns out to be an excellent indicator of the end of inflation. $\epsilon(t_i)$ is a small fraction of unity, but $\epsilon\uparrow1$ as $\boldsymbol\phi$ accelerates into the minimum at the origin. \change{This condition triggers before $\boldsymbol{\phi}$ oscillates around the minimum in the potential, so no information about those oscillations are stored. However, potential minima in Gaussian random fields are expected to resemble $N$-dimensional quadratics, so the oscillations may be easily approximated without further simulation.} Additionally, we define the ``inwardness'' parameter $\iota$ as the alignment between the slopes of $V$ and the minimum at any given location, or
\beq \label{eq:inwardness}
\iota(\boldsymbol\phi) \equiv (\hat\nabla_{\phi}V) \cdot \hat{\boldsymbol\phi}.
\eeq
where hats represent unit vectors. If $\iota<0$ at any point during the simulation, then $V$ is sloping away from the origin and it is extremely unlikely that $\boldsymbol\phi$ will converge to the $V=0$ minimum. \change{In the cases we examine below, we have verified that adopting this criterion results in rejecting a small fraction ($\lesssim 1\% $) of trajectories that do in fact converge to the correct minimum, and hence makes a negligible difference to the final probability distributions of observables.
However, as $N$ increases, there is a higher probability that $\boldsymbol{\phi}$ takes significant ($> 90^{\circ}$) turns before converging to the minimum, so we may have to relax this condition in future work.} As a last resort, $V$ itself is monitored. If $V<0$ at any point, then $\boldsymbol\phi$ is in the process of converging to a negative (and, therefore, unphysical) minimum away from the origin. All three conditions --- $\epsilon\uparrow1$, $\iota\downarrow0$, and $V\downarrow0$ --- serve to halt the simulate--solve cycle at the precise moment when either inflation ends or the evolution goes awry. The accumulated data is then bunched into an object and analyzed. Stored data includes the inflaton field, the potential, and stopping conditions at every location along the trajectory as well as the full covariance matrix.

\textit{Plausibility check.} Once the equations of motion are solved, \texttt{bushwhack} makes an effort to determine whether the solution is a remotely plausible representative of our own universe. The verifying questions that the code asks are as follows:
\begin{enumerate}
    \item Did the solution fail by diverging? In other words, was $\iota(\boldsymbol\phi)<0$ at any point? If so, do not analyze further. If not, proceed to the next check.
    \item Did the solution fail by submerging? In other words, was $V(\boldsymbol\phi)<0$ at any point? If so, do not analyze further. If not, proceed to the next check.
    \item Did the solution accumulate enough inflation? In other words, was $N_e(t_e) \geq 55$? If so, this solution is plausible enough and observable quantities can be extracted. If not, do not analyze further. 
\end{enumerate}
Even if the solution fails a plausibility check, the information is available for manual analysis.

\textit{Extract observables.} If a trajectory is found to converge to the origin with enough inflation, scalar, tensor, and isocurvature power spectra are extracted via methods detailed in Section \ref{sec:power_spectra}. To sample the \change{curvature} power spectrum, note from \eqref{eq:coupling_matrix} that it is now required to know the second derivatives of $V$ along the path, so those are simulated sparsely along the path using $\boldsymbol\Gamma$. It is worth emphasizing here that even with the same set of potential parameters and initial conditions on the trajectory, infinitely many random potentials and trajectories are possible, each of which yield different power spectra. It is only through repeated trials that we compile the \textit{statistical} predictions of observable quantities from a certain parameter set. Plots of data computed from one sample trajectory are presented in Figure \ref{fig:example_trajectory}.

\subsection{Algorithmic Data Compression} \label{sec:data_compression}

Even when $V$ is simulated only at points along the path evolution of $\boldsymbol\phi$, the covariance matrix $\boldsymbol\Gamma$ can become large and ill-conditioned. Since we add $N+1$ constraints with each time step, the covariance matrix $\boldsymbol\Gamma$ after $T$ time steps has over $(N+1)^2 T^2$ elements and accumulates over $(N+1)^2(2T+1)$ new elements each step. Both populating the matrix and performing operations with it, such as Cholesky decompositions, can quickly become burdensome, so compiling a representative sample of solutions with high $N$ without compression methods is computationally difficult. We present two methods of data compression that can be applied iteratively throughout a simulation to limit the size of $\boldsymbol\Gamma$ while maintaining reasonable levels of accuracy in predictions of spectral quantities.

\subsubsection{``Unprincipled'' Forgetting} \label{sec:unprincipled_forgetting}

The trajectory of $\boldsymbol\phi$ evolves smoothly down the slopes of $V$; at some time step $T$, the values nearest $\boldsymbol\phi_T$ are usually $\boldsymbol\phi_{T-1}, \boldsymbol\phi_{T-2}$, etc. By Equation \eqref{eq:covariance_matrix}, $V_T$ and its derivatives are thus most highly constrained by $V_{T-1}, V_{T-2}$, etc., and their derivatives. Similarly, if $\boldsymbol{\phi}$ travels long distances during its evolution, then $V_T$ is unlikely to be strongly constrained by early data. How much of the information contained in $\boldsymbol\Gamma$ contributes a non-negligible constraint on new data? It is possible that early, weak constraints can be deleted while retaining the recent, relevant core of $\boldsymbol\Gamma$ and therefore increasing the efficiency of solving for $\boldsymbol\phi$. We detail this forgetting method below, labeled ``unprincipled'' because the decision about which entries in $\boldsymbol{v}$ to forget is made in an \textit{ad hoc} manner.

\begin{figure}[t]
    \centering
    \includegraphics[width=3.4in]{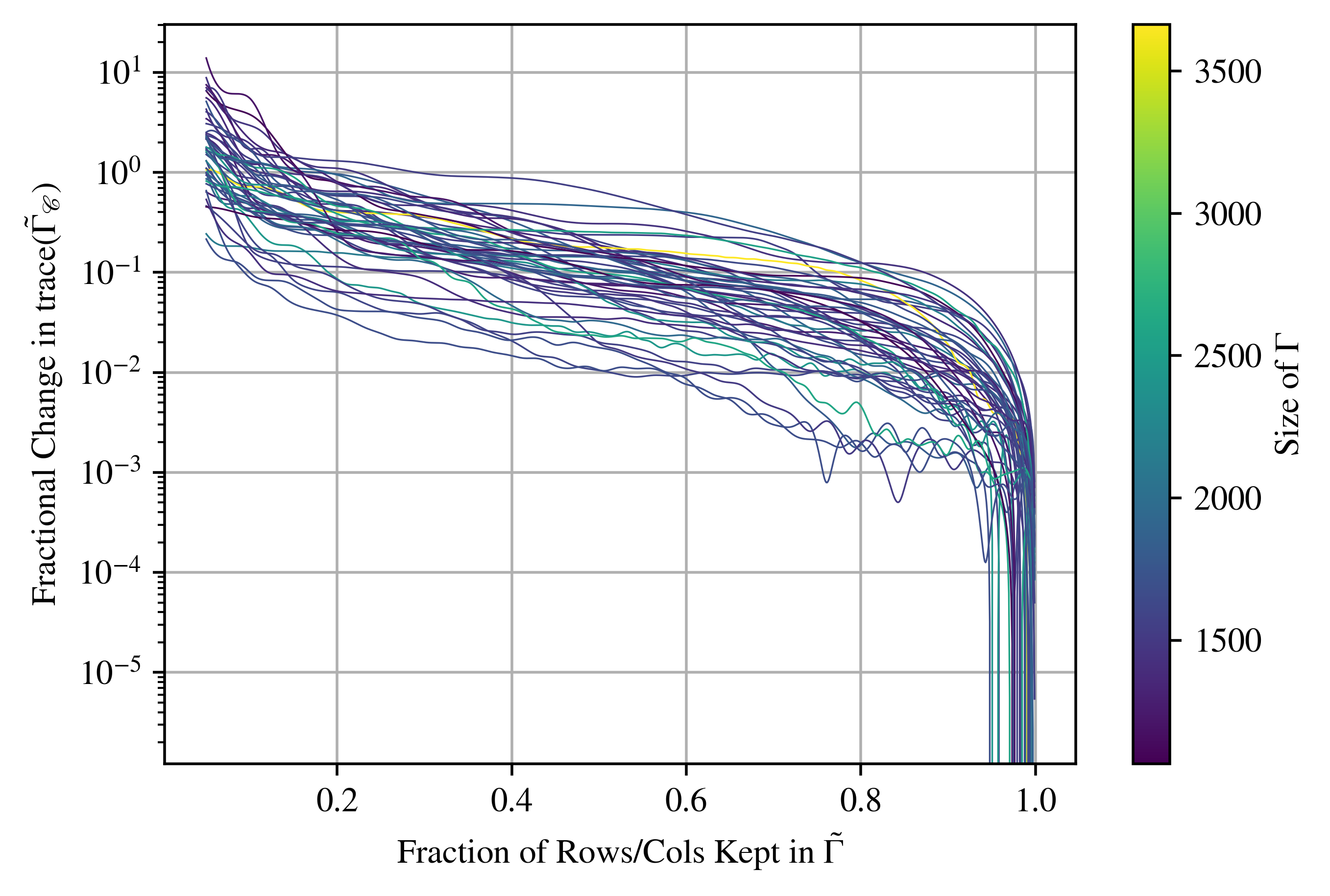}
    \caption{Constraining power lost from the covariance matrix $\boldsymbol\Gamma$ as a function of kept information. As more information is removed from earlier in the simulation (moving left), the constraints on a new potential value become more relaxed (the trace of the conditional covariance matrix $\tilde{\boldsymbol\Gamma}_{\mathcal{C}}$ increases). However, the constraining power of $\boldsymbol\Gamma$ is not equally distributed in its rows: half of all rows/columns can be removed before Tr$(\Tilde{\boldsymbol{\Gamma}}_{\mathcal{C}})$ increases by 10\%.}
    \label{fig:constraint_strength_vs_rows_kept}
\end{figure}

Assume that $\boldsymbol{\phi}$ is evolved through $T-1$ timesteps and that $\boldsymbol{\Gamma}$ is large and ill-conditioned. The strength of previous constraints on $V_T$ can be quantified in the sum of the eigenvalues (i.e., the trace) of $\boldsymbol{\Gamma}_{\mathcal{C}}$: \textit{lower} values of $\rm{Tr}(\boldsymbol{\Gamma}_{\mathcal{C}})$ correspond to \textit{stronger} constraints (lower variance in generated points). Consider removing rows/columns of $\boldsymbol\Gamma$ corresponding to early data, $V_i$, $V_1$, $V_2$ etc., one at a time, leaving a modified covariance matrix $\Tilde{\boldsymbol{\Gamma}}$, and recomputing the constrained covariance matrix $\Tilde{\boldsymbol\Gamma}_{\mathcal{C}}$ after each removal. We carried out this calculation with 50 converged trajectories in different random potentials. Figure \ref{fig:constraint_strength_vs_rows_kept} shows the strength of constraints from $\Tilde{\boldsymbol{\Gamma}}_{\mathcal{C}}$ relative to those from $\boldsymbol{\Gamma}_{\mathcal{C}}$ computed from the full covariance matrix as a function of the amount of data deleted. As $\Tilde{\boldsymbol{\Gamma}}$ gets smaller (moving from right to left on the plot), Tr$(\Tilde{\boldsymbol{\Gamma}}_{\mathcal{C}})$ increases (higher on the vertical axis). In most cases, half of the rows in $\boldsymbol\Gamma$ can be deleted before $\mathrm{Tr}(\Tilde{\boldsymbol\Gamma}_{\mathcal{C}})$ changes by more than $10\%$ of itself. However, the conditional covariance matrix does not directly tell us how deleting old data affects the accuracy of computed spectral quantities. 

\begin{figure*}
    \centering
    \includegraphics[width=7.0in]{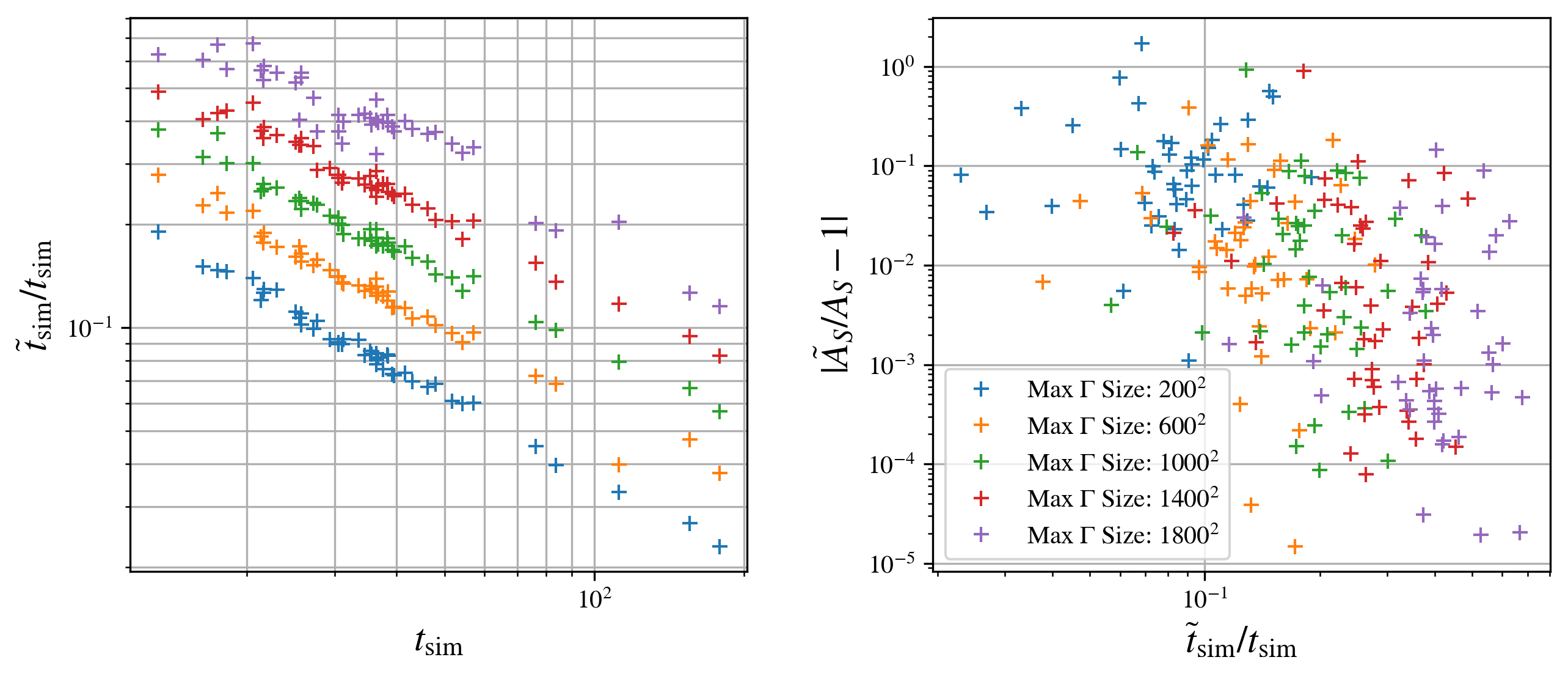}
    \caption{Quantifying compression and accuracy of ``unprincipled'' forgetting. \textit{Left}: A simulation without compression steps that takes a time $t_{\mathrm{sim}}$ [seconds] to complete can finish in a fraction of the time after incorporating forgetting steps. That fraction $\Tilde{t}_{\mathrm{sim}}/t_{\mathrm{sim}}$ falls as a power law with $t_{\mathrm{sim}}$. Colors represent the maximum allowed size of $\boldsymbol\Gamma$ before a compression step is executed: lower maximum size means more frequent compression. \textit{Right}: Even with high compression, simulations accurately reproduce uncompressed predictions. While higher compression is, indeed, less accurate, a moderate level of compression can be achieved while sustaining reasonably small error.} 
    \label{fig:unprincipled_forgetting_speedup_and_accuracy}
\end{figure*}

To jointly quantify the amount of time saved and the accuracy of the unprincipled forgetting method, we ran 50 simulations with the \texttt{bushwhack} code, parameters $(s, V_{\star}^2, N, \phi_i) = (30, 5\times10^{-9}, 2, 0.8s)$, without any compression methods. For each run, we saved the total simulation time $t_{\mathrm{sim}}$, predicted value for $A_S$, and random state, then reran the simulation (with the same random state) 5 times, each executing unprincipled forgetting steps once $\boldsymbol\Gamma$ reached some variable maximum size. For each rerun we, again, stored the compressed simulation time $\Tilde{t}_{\mathrm{sim}}$ and compressed prediction for the \change{curvature} spectral amplitude $\Tilde{A}_S$. The left panel of Figure \ref{fig:unprincipled_forgetting_speedup_and_accuracy} demonstrates that the fractional time taken $\Tilde{t}_{\mathrm{sim}}/t_{\mathrm{sim}}$ decreases as a power law with $t_{sim}$. The time saved can achieve values of $1/5$ with conservative compression and $1/50$ with high compression while simultaneously predicting $\Tilde{A}_S$ to be very nearly correct. The right panel of Figure \ref{fig:unprincipled_forgetting_speedup_and_accuracy} confirms that higher degrees of compression tend to produce less accurate predictions, but assures that these inaccuracies are small enough at moderately high levels. Given the speed-up, accuracy, and ease of implementation of unprincipled forgetting steps, we incorporate them permanently into the \texttt{bushwhack} code to execute at a maximum $\boldsymbol\Gamma$ size of $1000^2$.

\subsubsection{``Principled'' Forgetting} \label{sec:principled_forgetting}

In the previous subsection, we described an method for increasing computational efficiency that involved compressing $\boldsymbol\Gamma$ under the assumption that earlier points weakly constrain newly generated data. This method was labeled ``unprincipled'' because it involved the reasonable, but \textit{ad hoc} choice to delete points based on their distance from the point of interest. In this section, we describe a ``principled'' method of compressing $\boldsymbol\Gamma$ that allows for direct control over the portion of trace kept. We leave its implementation in random field inflationary trajectories to future work (see Section \ref{sec:future_work}).

We seek to compress $\boldsymbol\Gamma$ so as to maximize its constraining power while minimizing its overall size. When $\boldsymbol{v}_{\mathcal{P}}$ gets large, we seek to find a compression matrix $\boldsymbol{A}$ such that
\beq \label{eq:compression_matrix}
\Tilde{\boldsymbol{v}}_{\mathcal{P}} = \boldsymbol{A}\boldsymbol{v}_{\mathcal{P}},
\eeq
where $\Tilde{\boldsymbol{v}}_{\mathcal{P}}$ is a smaller, compressed version of $\boldsymbol{v}_{\mathcal{P}}$. We can transform blocks of $\boldsymbol\Gamma$ into compressed forms as
\begin{align}
    \Tilde{\boldsymbol\Gamma}_{\mathcal{PP}} &= \langle \Tilde{\boldsymbol{v}}_{\mathcal{P}} \Tilde{\boldsymbol{v}}_{\mathcal{P}}^T \rangle = \boldsymbol{A}\boldsymbol\Gamma_{\mathcal{PP}}\boldsymbol{A}^T \label{eq:Gamma_PP_tilde}\\
    \Tilde{\boldsymbol\Gamma}_{\mathcal{FP}} &= \langle \Tilde{\boldsymbol{v}}_{\mathcal{F}} \Tilde{\boldsymbol{v}}_{\mathcal{P}}^T \rangle = \boldsymbol\Gamma_{\mathcal{FP}}\boldsymbol{A}^T \label{eq:Gamma_FP_tilde}\\
    \Tilde{\boldsymbol\Gamma}_{\mathcal{C}} &= \boldsymbol\Gamma_{\mathcal{FF}} - \Tilde{\boldsymbol\Gamma}_{\mathcal{FP}}\Tilde{\boldsymbol\Gamma}_{\mathcal{PP}}^{-1}\Tilde{\boldsymbol\Gamma}_{\mathcal{PF}}. \label{eq:Gamma_C_tilde}
\end{align}
As mentioned previously, \textit{lower} values of $\mathrm{Tr}(\boldsymbol\Gamma_{\mathcal{C}})$ correspond with \textit{tighter} constraints on new potential generation. Thus, to maximize the constraining power of $\Tilde{\boldsymbol\Gamma}$, we need to reduce $\mathrm{Tr}(\Tilde{\boldsymbol\Gamma}_{\mathcal{C}})$ by maximizing the trace of its latter term,
\begin{align}
\tau &\equiv \mathrm{Tr}(\Tilde{\boldsymbol\Gamma}_{\mathcal{FP}}\Tilde{\boldsymbol\Gamma}_{\mathcal{PP}}^{-1}\Tilde{\boldsymbol\Gamma}_{\mathcal{PF}})\\
&= \mathrm{Tr}(\boldsymbol\Gamma_{\mathcal{FP}}\boldsymbol{A}^T(\boldsymbol{A}\boldsymbol\Gamma_{\mathcal{PP}}\boldsymbol{A}^T)^{-1}\boldsymbol{A}\boldsymbol\Gamma_{\mathcal{PF}}) \label{eq:tau}
\end{align}
It is straightforward to show that $\tau$ only depends on the row space of $\boldsymbol{A}$, so we are free to impose the additional constraint that the rows of $\boldsymbol{A}$ be orthonormal with respect to $\boldsymbol\Gamma_{\mathcal{PP}}$.
\beq \label{eq:orthonormality}
\boldsymbol{A}\boldsymbol\Gamma_{\mathcal{PP}}\boldsymbol{A}^T = \boldsymbol{1}
\eeq
This constrained maximization problem leads to
\beq \label{eq:constrained_maximization_problem}
\boldsymbol{A}\boldsymbol\Gamma_{\mathcal{FP}}\boldsymbol\Gamma_{\mathcal{PF}} = \boldsymbol\Lambda\boldsymbol{A}\boldsymbol\Gamma_{\mathcal{PP}}
\eeq
where $\boldsymbol\Lambda$ is a matrix of Lagrange multipliers. We can once again take linear combinations of the rows of $\boldsymbol{A}$ to diagonalize $\boldsymbol\Lambda$. Now, each row $\boldsymbol{a}$ of $\boldsymbol{A}$ is a solution to the generalized eigenvalue problem
\beq \label{eq:A_eigenvalue_problem}
\boldsymbol\Gamma_{\mathcal{FP}}\boldsymbol\Gamma_{\mathcal{PF}}\boldsymbol{a} = \lambda\boldsymbol\Gamma_{\mathcal{PP}}\boldsymbol{a}.
\eeq
Since $\tau = \sum_i \lambda_i$, the optimal compression matrix will comprise some number of rows corresponding to the largest eigenvalues so that the size of $\Tilde{\boldsymbol\Gamma}_{\mathcal{PP}}$ is much less than $\boldsymbol\Gamma_{\mathcal{PP}}$ and $\tau \approx \mathrm{Tr}(\boldsymbol\Gamma_{\mathcal{FP}}\boldsymbol\Gamma_{\mathcal{PP}}^{-1}\boldsymbol\Gamma_{\mathcal{PF}})$. Significant compression may be possible with this method, as the density of data points required to resolve the path evolution of $\boldsymbol\phi$ is often greater than what is necessary to describe the potential in a local domain. An overdensity of data points leads to many redundant constraints on new potential generation, and thus a high concentration of $\tau$ in the very largest eigenvalues. We leave proper implementation of principled forgetting to future work.

\section{Results} \label{sec:results}

\begin{figure}[b]
    \centering
    \includegraphics[width=3.4in]{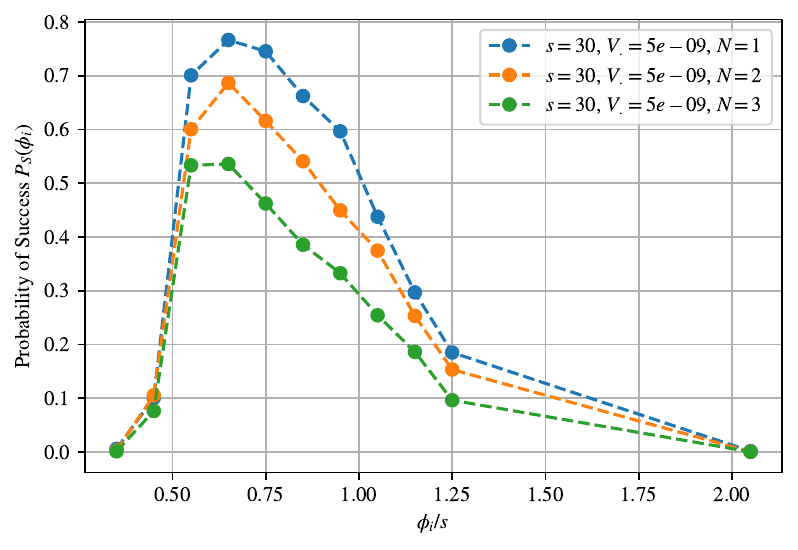}
    \caption{The probability that a trajectory starting at an initial distance $\phi_i$ from the origin will both converge and accumulate enough inflation. This function is necessary when computing the weighted mean prediction for any spectral quantity.}
    \label{fig:success_probabilities}
\end{figure}

\begin{table*}
\centering
\begin{tabular}{|p{0.75in}|p{0.6in}|p{0.6in}|p{0.6in}|p{0.6in}|p{0.6in}|p{0.6in}|p{0.6in}|p{0.6in}|p{0.6in}|}
\hline
\textbf{Spectral\newline Quantity} & $\mu_q\newline (N=1)$ & $\sigma_q\newline (N=1)$ & Tension\newline $(N=1)$ & $\mu_q\newline (N=2)$ & $\sigma_q\newline (N=2)$ & Tension\newline $(N=2)$ & $\mu_q\newline (N=3)$ & $\sigma_q\newline (N=3)$ & Tension\newline $(N=3)$\\
\hline
\hline
$\log_{10}(A_S)$ & $-9.460$ & $0.054$ & $14.4\sigma$ & $-9.465$ & $0.080$ & $9.8\sigma$ & $-9.463$ & $0.057$ & $13.6\sigma$\\
\hline
$n_S$ &$0.9637$ & $0.0019$ & $0.2\sigma$ & $0.9613$ & $0.0022$ & $0.2\sigma$ & $0.9598$ & $0.0017$ & $0.5\sigma$\\
\hline
$\log_{10}(A_T)$ & $-10.392$ & $0.096$ & N/A & $-10.420$ & $0.070$ & N/A & $-10.430$ & $0.042$ & N/A\\
\hline
$n_T$ & $-0.0158$ & $0.0019$ & N/A & $-0.0171$ & $0.0009$ & N/A & $-0.0178$ & $0.0005$ & N/A\\
\hline
$\log_{10}(A_{\rm{iso}})$ & N/A & N/A & N/A & $-13.792$ & $0.123$ & N/A & $-13.166$ & $0.070$ & N/A\\
\hline
$n_{\rm{iso}}$ & N/A & N/A & N/A & $0.9688$ & $0.0023$ & N/A & $0.9659$ & $0.0017$ & N/A\\
\hline
$\log_{10}(r)$ & $-0.907$ & $0.064$ & $1.4\sigma$ & $-0.929$ & $0.059$ & $1.2\sigma$ & $-0.940$ & $0.047$ & $1.3\sigma$\\
\hline
$\log_{10}(\beta_{\rm{iso}})$ & N/A & N/A & N/A & $-4.325$ & $0.059$ & 0 & $-3.704$ & $0.051$ & 0\\
\hline
\end{tabular}
\caption{Tabulated results and comparisons. Shown are the weighted means, standard deviations, and tensions with Planck data of various spectral quantities for each parameter set \eqref{parameter_set_1}-\eqref{parameter_set_3}. While $A_S$ predictions exhibit significant tension with Planck data, slight variations in $s$ and/or $V_{\star}$ may remedy these differences while preserving the agreements seen in $n_S$, $r$, and $\beta_{\rm{iso}}$.}
\label{table:results}
\end{table*}

The \texttt{bushwhack} code (see Section \ref{sec:program}) was parallelized and implemented with unprincipled forgetting (see Section \ref{sec:unprincipled_forgetting}) for execution on the University of Richmond \texttt{Quark} Supercomputing Cluster. As a preliminary investigation, we choose some reasonable parameter sets for the generation of $V$:
\begin{align}
    (s, V_{\star}^2, N) &= (30, 5 \times 10^{-9}, 1) \label{parameter_set_1}\\
    (s, V_{\star}^2, N) &= (30, 5 \times 10^{-9}, 2) \label{parameter_set_2}\\
    (s, V_{\star}^2, N) &= (30, 5 \times 10^{-9}, 3) \label{parameter_set_3}
\end{align}
The only variable between these sets is $N$, which may change predictions of observable quantities in nontrivial ways, though varying $s$ and $V_{\star}$ is worthwhile future work.

To gather a representative sample of trajectories from a given parameter set, we catalogue $1000$ simulations with $\phi_i$ chosen from each of $12$ radial $\phi$-shells of thickness $0.1s$ from $0.3s$ to $2.1s$. That is, we choose the initial distance from the minimum to lie within a certain bin (for instance, $\phi_i \in [0.8s,0.9s]$), then randomly choose the vector location $\boldsymbol\phi_i$, all $1000$ times in each bin. If $\phi_i < 0.3s$, the trajectory is very unlikely to accumulate enough inflation, and if $\phi_i > 2.1s$, exponentially unlikely to converge to the origin.

\subsection{Weighting Predictions by Success Probabilities}
\label{sec:weighting}

We seek statistical predictions for observable quantities for every $s,V_{\star},N$, and $\phi_i$. The simplest way to collect a representative sample of successful trajectories would be to choose $\boldsymbol\phi_i$ uniformly in the volume of a large sphere (say, with radius $3s$) centered on the minimum at the origin, many times for many random potentials, and save only the successful runs. This would naturally represent all possible successful trajectories and the statistical predictions could be calculated by simple means and variances. However, in this approach, most random $\phi_i$ would be large -- toward the outskirts of the sphere -- because most of the sphere's volume is at large radii. To obtain a sizeable number of successful trajectories with low $\phi_i$, the number of total solutions would need to be very high, and thus computationally expensive, especially in high-$N$ potentials.

Instead, we subdivide the $\phi$-volume into spherical shells of equal thickness (not equal volume) around the origin and catalog the same number of trajectories with $\phi_i$ in each shell. This allows us to store an equal number of trajectories with $\phi_i \approx 0.4s$ as with $\phi_i \approx 2.0s$. For instance, if parameter set \eqref{parameter_set_2} yields especially interesting predictions in the $\phi_i\in [0.3s,0.4s]$ bin, this method ensures there are plenty of simulations with this property to examine. However, when computing overall predictions for spectral quantities associated with a particular $(s,V_{\star},N)$ parameter set, the collection of trajectories from all shells is not representative of all possible successful trajectories, since we have sampled more densely at small $\phi_i$. So, we must weight the means and variances computed from each shell accordingly.

If $q_l$ is the simple mean of predictions for a spectral quantity $q$ from trajectories with $\phi_i$ in shell $l$, then the weighted mean $\mu_q$ considers all shells as
\beq \label{eq:weighted_mean}
\mu_q = \frac{\sum_l w_l q_l}{\sum_l w_l}
\eeq
where $w_l$ is a shell weight. The weights are the probability that a random successful trajectory from a given parameter set has $\phi_i \in l$. We can compute this probability density function by multiplying the probability $P_S(\phi_i)$ that a trajectory succeeds -- converges to the origin with enough inflation -- with a volume element to account for more volume in shells farther from the $\phi$-origin. $P_S$ is an interesting function of $\phi_i$, plotted in Figure \ref{fig:success_probabilities}, and is computed by taking the simple ratio of successful trajectories to total attempted runs. The probability that a trajectory is successful vanishes in both $\phi_i\downarrow0$ and $\phi_i\uparrow\infty$ limits (because of $e$-fold and convergence requirements, respectively) and peaks around $\phi_i = 0.7s$ for each considered parameter set. With this probability evaluated in each shell, we interpolate linearly for a smooth PDF and find the weights to be
\begin{equation}
\label{eq:weights}
    w_l = \frac{1}{\rm{norm}}\int_l P_S(\phi_i) \mathrm{SA}_N(\phi_i) d\phi_i
\end{equation}
where $\mathrm{SA}_N(\phi_i) \propto R^{N-1}$ is the surface area of an $N$-sphere of radius $\phi_i$. Similar to $\mu_q$, 
the variance of $q$ is
\beq \label{weighted_variance_def}
\sigma_q^2 = \frac{\sum_l w_l q_l^2}{\sum_l w_l} - \mu_q^2.
\eeq

\subsection{Power Spectra Predictions}

\begin{figure*}
    \centerline{
    \includegraphics[trim={0 5cm 0 6cm}, width=8.0in]{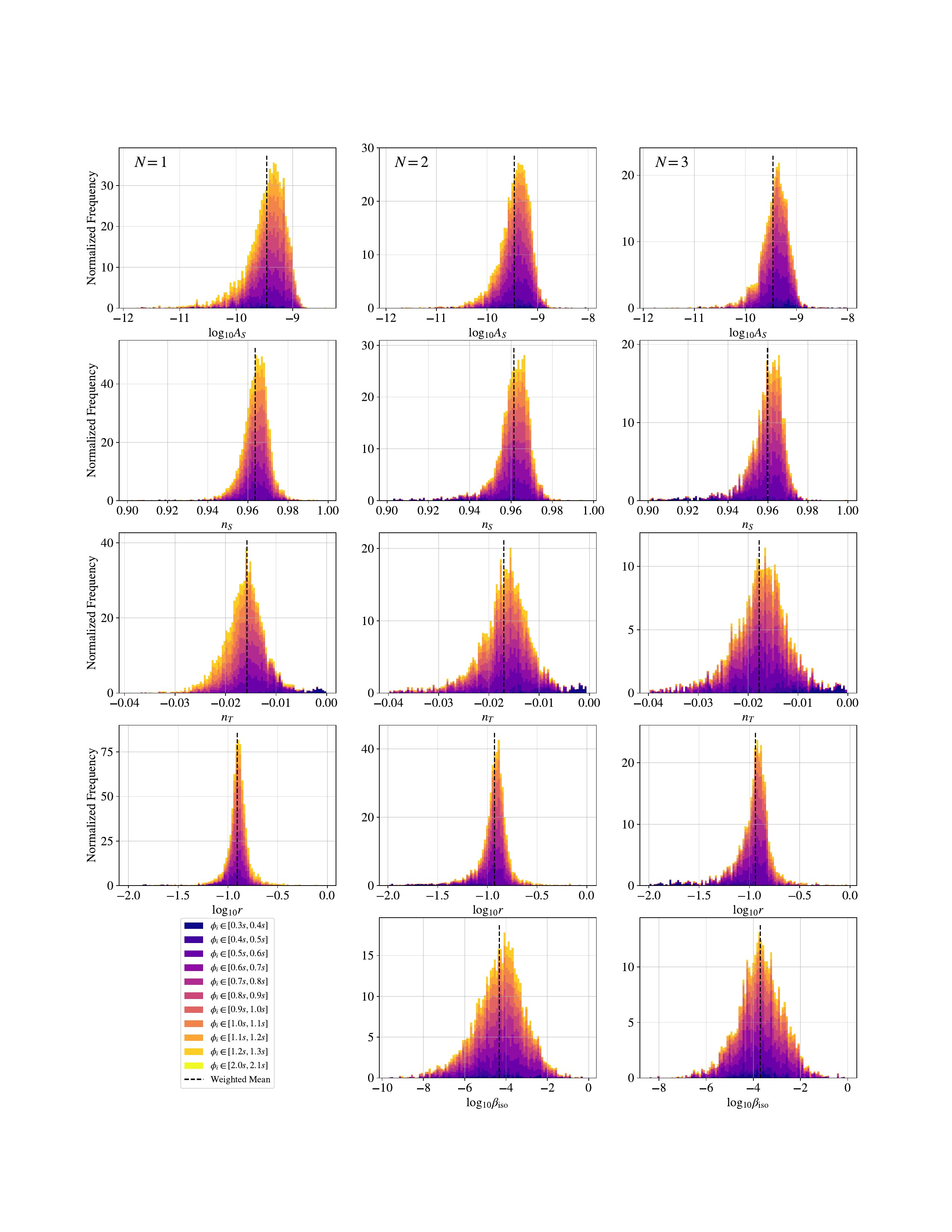}
    }
    \caption{Weighted histograms of observable quantities $A_S$, $n_S$, $n_T$, $r$, and $\beta_{\rm{iso}}$. Different colors on each histogram correspond to different initial positions $\phi_i$. Each row of plots corresponds to a particular spectral quantity and each column to a parameter set. Horizontal axes are the same along each row.}
    \label{fig:histograms}
\end{figure*}

We now present the statistical predictions for observable quantities computed from a Gaussian random inflaton potential with parameters sets \eqref{parameter_set_1}--\eqref{parameter_set_3}. We assume any error resulting from compression methods is accurately represented in the variance of the outputted spectral quantities. First, we examine Figure \ref{fig:histograms}, which shows weighted, normalized histograms for each of $A_S, n_S, n_T, r$, and $\beta_{\rm{iso}}$ for each parameter set, colored by initial condition. A few observations to note:
\begin{itemize}
    \item Changing $N$ does not significantly affect the mean or variance of any quantity, but weak trends are observed in $n_S$ (positive), $A_T$ (negative), $n_T$ (negative), and $r$ (negative). Future work into higher $N$ is required to confirm the sensitivity of perturbation spectra to the dimension of $V$.
    \item Distributions within individual $\phi_i$ shells are not necessarily the same as the combined distribution for the parameter set. This is especially apparent in the low $\phi_i$ regime (darker blues); simulations with low $\phi_i$ predict values of $n_S$, $n_T$, and $r$ that are significantly offset from the overall mean (which is dominated by simulations with intermediate $\phi_i$).
    \item Predicted mean values are in variable tension with Planck observations, detailed in Table \ref{table:results}. For example, $A_S$ is lower than the Planck prediction for all $N=1,2,3$, while $n_S$, $r$, and $\beta_{\rm{iso}}$ predictions are accommodated by all three parameter sets. Note that this does not indicate that the Gaussian random potential is unable to accommodate all Planck predictions, but that the particular parameter sets \eqref{parameter_set_1}--\eqref{parameter_set_3} are not the best fit.
    \item Distributions roughly normal, but some have slight negative skewness (in particular $A_S$ and $n_S$). Thus, the variance may not be the best metric of statistical spread for each quantity. 
\end{itemize}
The predictions visualized in each histogram are quantified in Table \ref{table:results}. We also present a corner plot of these $5$ spectral quantities for our $N=3$ simulations. These panels further emphasize a few points:
\begin{itemize}
    \item Most of the quantities appear to be uncorrelated, however weak correlations are observed in $n_T$ vs. $n_S$ (positive) as well as $n_T$ vs. $r$ (negative). Some of the weighted KDEs exhibit curving, non-trivial contours. 
    \item Weighting by Equation \eqref{eq:weights} exaggerates the contributions from simulations with $\phi_i \in [0.5s,1.2s]$ and minimizes contributions from those outside this range, leaving tight kernel density estimations around the means.
\end{itemize}

\begin{figure*}[p]
    \centerline{
    \includegraphics[width=7.0in]{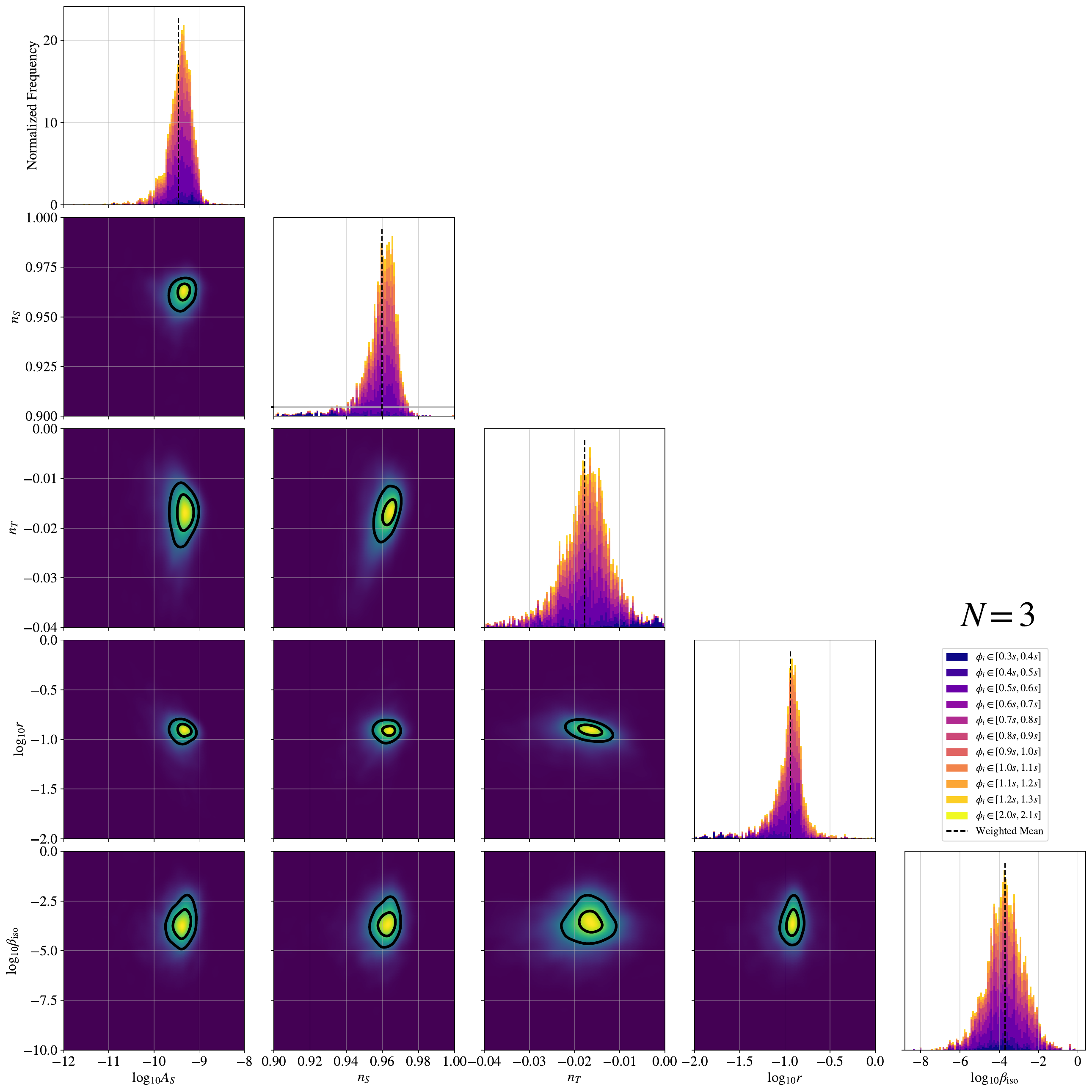}
    }
    \caption{Corner plot of histograms and Gaussian kernel density estimations (KDEs) of correlations between spectral quantities for the $N = 3$ parameter set. On the diagonal, histograms are identical to those plotted in the right column of Figure \ref{fig:histograms}. On the off-diagonals, KDEs are computed from weighted scatterplots of $\sim 10^4$ simulations: lighter regions indicate the most frequently outputted pairs of parameters while darker regions are uncommon outputs. Two closed black curves are included with each KDE to clarify the subtleties in its shape. The spectral quantities $n_S$, $n_T$, $r$, and $\beta_{\rm{iso}}$ are shown to be independent of each other or weakly correlated, at best, for this parameter set.}
    \label{fig:corner_plot}
\end{figure*}

\section{Conclusions} \label{sec:conclusions}

We have presented a method for exploration into $N$-dimensional inflationary models in which the inflaton potential is a Gaussian random field. We employ several techniques to increase computational efficiency and suggest another for implementation in future work. To be specific, we avoid generating the random potential all at once, but efficiently generate along the path evolution of the inflaton field. The covariance matrix of constraints $\boldsymbol\Gamma$ can grow large and unwieldy, especially with large $N$, so we propose two compression methods that may ease the computational intensity by exploiting redundant constraints within $\boldsymbol\Gamma$. We implement ``unprincipled forgetting'' steps---deleting old constraints to shrink $\boldsymbol{\Gamma}$---into our solver and compile thousands of simulations to compute statistical predictions of \change{curvature}, tensor, and isocurvature power spectra. We list our main conclusions below:
\begin{itemize}
    \item The Gaussian random potential is a highly versatile inflationary model with a rich volume of promising parameter space. Parameters $N$ and $\phi_i$ each change the statistical predictions for perturbation spectra in unique ways. For example, higher $N$ tends to increase $n_S$ while decreasing $n_T$, $A_T$, and $r$ and lower $\phi_i$ predicts lower $n_S$ and higher $n_T$. Parameter sets \eqref{parameter_set_1}--\eqref{parameter_set_3} have been demonstrated to accommodate modern constraints on $n_S$, $r$, and $\beta_{\rm{iso}}$, but not $A_S$ (see Table \ref{table:results} and Figures \ref{fig:histograms}, \ref{fig:corner_plot}).
    \item Methods such as the one we describe allow for broad exploration of the space of possible outcomes in any given model. In particular, we do not restrict our attention in advance to trajectories that have certain desired properties; rather, we explore the space of all possible trajectories (subject only to the anthropic constraint that inflation end). This is the right approach if one wants to make probabilistic statements about the probability distribution of any given set of observables in a given model.
    \item Algorithmic data compression (Section \ref{sec:principled_forgetting}) allows for the possibility of significantly more efficient generation of $V$ in high-$N$ regimes. Unprincipled forgetting is limited in that it relies on the assumption that earlier data contributes weaker constraints on new generation and less rigorous in that it involves the deletion of data rather than compression. Principled forgetting methods, if implemented efficiently, could remedy these issues.
\end{itemize}

The examples we have presented in this paper are intended as illustrations of the method and are far from exhaustive. For example, the models we consider always involve large excursions (significantly greater than the Planck mass) in field space. It is possible that using a fixed effective field theory over such large excursions is incorrect; in any case, it is of interest to explore other parameter choices. Similarly, we restrict our attention to the case of a flat field-space metric, but our method could be extended to accommodate any given non-flat metric. Most importantly, it is of course of great interest to explore higher values of $N$, which will be possible with a combination of improvements in efficiency and simply throwing more CPU cycles at the problem.

\subsection{Future Work} \label{sec:future_work}

This project could be further explored in a number of ways, ranked below by priority.
\begin{enumerate}
    \item \textit{Explore parameter space.} Carry out a large number of simulations with different parameter sets which vary $s$ and $V_{\star}$ in addition to $N$. These supplemental runs could quantify the degree to which the Gaussian random potential model is able to accurately reproduce modern constraints on primordial power spectra.
    \item \textit{Implement principled forgetting.} Unprincipled forgetting offers significant speed increases, but at the expense of rigorous, algorithmic data compression. Principled forgetting methods detailed in Section \ref{sec:principled_forgetting} have the capacity to provide enormous compression to $\boldsymbol\Gamma$, possibly enough to efficiently explore string-theory-esque models with $N \sim 100$.
    \item \textit{Incorporate other observables.} The spectral quantities computed from our trajectories are but a subset of the independent quantities constrained by modern observations. Incorporating quantities such as the non-Gaussianity of the CMB could further verify whether the Gaussian random potential is a plausible inflationary model of our universe \citep{renaux-patel}. \change{Computing non-Gaussianity may require modifying our stopping criterion to include oscillations around the potential minimum.}
    \item \textit{Translate to other languages.} The \texttt{bushwhack} code is currently written in Python, but could be optimized by translating to more efficient languages such as C or FORTRAN.
\end{enumerate}

\section*{Acknowledgements}

This project was supported by two Summer Research Fellowships from the University of Richmond School of Arts \& Sciences from May 2019 to July 2020. Simulations were run on the University of Richmond \texttt{Quark} Supercomputing Cluster, purchased with a Major Research Instrumentation grant from the National Science Foundation. EFB thanks Andrew Jaffe and Alan Heavens for enlightening conversations and Imperial College for its hospitality during the beginning of this work.

\bibliographystyle{mnras}
\bibliography{Bibliography}

\end{document}